\documentstyle[11pt]{article}
\begin{document}  

\renewcommand{\thefootnote}{\fnsymbol{footnote}}
\renewcommand{\theequation}{\arabic{section}.\arabic{equation}}  

\parskip = 0 pt  
\parindent = 10 pt  
\abovedisplayskip=13pt plus 3pt minus 9pt
\belowdisplayskip=13pt plus 3pt minus 9pt
  
\centerline{\LARGE \bf Subdynamics as a mechanism for objective
description}\par
\vskip 25 pt
\centerline{Ludovico~Lanz$^{{(1)}}$, Olaf~Melsheimer$^{(2)}$ and
Bassano~Vacchini$^{(1,2)}$}
\vskip 5 pt
{\it  
\centerline{$^{{(1)}}$
Dipartimento di Fisica dell'Universit\`a di
Milano and INFN, Sezione di Milano,}
\centerline{Via Celoria 16,  
I-20133, Milano, Italy}  
\centerline{$^{(2)}$Fachbereich Physik, Philipps--Universit\"at,
Renthof 7, D-35032, Marburg, Germany}
}  
\vskip 25 pt
\centerline{\sc Abstract}\par  
{ \baselineskip=12pt 
\textit{The relationship between microsystems and macrosystems is
considered
in the context of quantum field formulation of statistical mechanics:
it is argued that problems on foundations of quantum mechanics can be
solved relying on this relationship. This discussion requires some
improvement of non-equilibrium
statistical mechanics that is briefly presented.}
}\par  
\vskip 25 pt
\noindent  
\par  
\section{Evolution of the atomistic idea: the new
relationship micro-macrosystems}
\label{lucrezio}
\par         
Classical physics was fully compatible with a deeply rooted,
venerable philosophy: the wealth of objects called
{\it macrosystems},
we become aware of in our experience at a certain space-time
scale, arises by different structures of very many, very
elementary, possibly of very few kinds, unavoidably elusive
objects, 
{\it microsystems} at the previous scale; 
already in ancient times Lucretio called them {\it
primordia rerum}. 
For a physicist the fundamental challenge
is to learn about microsystems and discover the basic links;
what remains is to face complexity: correlations of simple
properties of single microsystems provide all describable
variety. One expects that this scheme can be reproduced at
smaller time-scales, to leave open the adventure of
discerning more remote links. Since generally human being
feels himself not completely fitting inside the
forementioned scheme, one might be happy to cut himself
apart from the world of objects, assuming the role of an
{\it observer} and, nice enough, quantum mechanics seemed to
support this dream. Belief in all this is rather
popular, has however a drawback: it makes typical  quantum
mechanical features like entanglement between microsystems
paradoxical and puzzling. Therefore much effort has been
done to experimentally check these typical features, which
are indeed deeply rooted into the basic
mathematical structure of quantum mechanics: as a result quantum
mechanics has passed these tests in a brilliant way.
Now a more subtle philosophy should be assumed, which as
always happens in physics does not mean complete dismission
of the previous one, but a reappraisal of it: something of
it remains true, e.g., when some classical limit holds; the
basic ideas however should change.
As it is well-known quantum mechanics is equivalent to a
quantum field theory based on quantization of classical
interacting and self-interacting Schr\"odinger fields (one
for each type of particle); in the case of
macrosystems, as it will be shown in \S~\ref{torun}, the field
formulation becomes very important for the construction of
the densities of conserved quantities and the related
currents; finally decisive is the fact that the field
approach becomes mandatory if relativity is taken into
account.
The atomistic model essentially transforms into the picture
of a dynamics for any  system, as complex as it will be, that
is driven by local, simple, universal interactions between
fields. Quantization on the other hand implies that mass and electric
charges and other conserved quantities have a discrete
structure, thus pointing to the existence of {\it particles}
that acquire a leading role in any irreducible {\it event} produced by the
field dynamics.
We would like to call attention to a crucial point, that
is taken in due consideration in quantum
optics but is generally pushed back,
probably due to the old philosophy,
in quantum field theory: the
macroscopic setting which characterizes any concrete system
implies boundary conditions on the fields by which {\it normal
modes} arise. Mass, charge and energy of the system are then
distributed through quantum population of this huge set of
normal modes which are in turn involved if these
populations are changed due to local interactions. In this
way an objective macroscopic property of the physical
system, precisely its {\it shape}, 
influences in an important way what the microsystems
leading the quantum dynamics are, at complete variance
with the old philosophy. Take for example massive matter inside a
container, with perfectly reflecting walls, enclosing a
space region $\omega$; normal modes are then a set of stationary
waves, complete and orthonormal in
$L^2 (\omega)\otimes {\mbox{\bf C}}^{2s+1}$,
$u_r({\mbox{\bf x}},\sigma)e^{-{i\over \hbar}W_r
t}$, $\sigma$ denoting the spin variable, such that
        \[  
        -{  
        \hbar^2  
        \over  
        2m  
        }  
        \Delta_2  
        u_r({\mbox{\bf x}},\sigma)= W_r
        u_r({\mbox{\bf x}},\sigma)
        \qquad  
        u_r({\mbox{\bf x}},\sigma)=0 \quad {\mbox{\bf x}}\in
        \partial\omega  
        .
        \]  
The Schr\"odinger  operator is given by
        \[
        {\hat \psi}({\mbox{\bf x}},\sigma)
        =
        \sum_r
        {\hat a}_r
        u_r({\mbox{\bf x}},\sigma),
        \]
with ${
        [  
        {\hat a}_{ r},
        {\hat a}_{r'}^{\scriptscriptstyle\dagger}
        ]  
        }_\pm  
        =  
        \delta_{r,r'}
       $,
the basic quantity is mass
$
        {\hat M}
        =
        \int_{\omega} d^3\!
        {\mbox{\bf x}}
        \,
        {\hat {\rho}}({\mbox{\bf x}})
=
        m
        \sum_r
        {\hat a}_{r}^{\scriptscriptstyle\dagger}
        {\hat a}_r
$, where the density is
$
        {\hat {\rho}}({\mbox{\bf x}})=
        m
        \sum_{\sigma}
         {\hat \psi}^{\scriptscriptstyle\dagger}
        ({\mbox{\bf x}},\sigma)
         {\hat \psi}
        ({\mbox{\bf x}},\sigma)
$. 
More refined representations of ${\hat M}$ in terms of a phase-space
density occur in kinetic theory.
The Hamiltonian is
        \[
        {\hat H}=
        \sum_{\sigma}
        \int_{\omega} d^3\!
        {\mbox{\bf x}}
        \,
        {\hat e}({\mbox{\bf x}},\sigma)
        \]
with
        \begin{eqnarray}
        \label{1.1}
        {\hat e}({\mbox{\bf x}},\sigma)
        &=&
        {{  
        \hbar^2  
        \over  
               2m  
        }}  
        \nabla
        {\hat \psi}^{\scriptscriptstyle\dagger}({\mbox{\bf x}},\sigma)
        \cdot  
        \nabla
        {\hat \psi}({\mbox{\bf x}},\sigma)
        +  
        {\hat \psi}^{\scriptscriptstyle\dagger}
        ({\mbox{\bf x}},\sigma)
        U({\mbox{\bf x}},t)
         {\hat \psi}
        ({\mbox{\bf x}},\sigma),
        \\
        &\hphantom{=}&
        {}
        +
        {1\over 2}  
        \sum_{\sigma'}
        \int_\omega d^3\! {\mbox{\bf  y}}
        \,  
         {\hat \psi}^{\scriptscriptstyle\dagger}
        ({\mbox{\bf x}},\sigma)
         {\hat \psi}^{\scriptscriptstyle\dagger}
        ({\mbox{\bf y}},\sigma')
        V(
        \left |
        {\mbox{\bf x}} - {\mbox{\bf y}}
        \right |
        )
         {\hat \psi}
        ({\mbox{\bf y}},\sigma')
         {\hat \psi}
        ({\mbox{\bf x}},\sigma),
        \nonumber
        \end{eqnarray}
where for simplicity only one  massive
self-interacting field has been considered and
$
        V(
        \left |
        {\bf x} - {\bf y}
        \right |
        )
$
describes the short range (mass) density-density interaction, while 
$
U({\mbox{\bf x}},t)
$ describes an external field affecting all systems constituted by the
field 
${\hat \psi}({\mbox{\bf x}},\sigma)$ 
(e.g., a gravitational field).
Also momentum is an important quantity
\begin{displaymath}
        {\hat {\mbox{\bf p}}}
        =  
        \frac {1}{2}     \!  
        \sum_{\sigma}
        \int_{\omega} d^3\!
        {\mbox{\bf x}}
        \,
        \left \{  
        \left[
        i\hbar
        \nabla
        {\hat
        \psi}^{\scriptscriptstyle\dagger}({\mbox{\bf
        x}},\sigma)
        \right]  
        {\hat \psi}({\mbox{\bf x}},\sigma)
        -
        {\hat
        \psi}^{\scriptscriptstyle\dagger}({\mbox{\bf
        x}},\sigma)
        i\hbar
        \nabla
        {\hat \psi}({\mbox{\bf x}},\sigma)
        \right \}. 
\end{displaymath}
This
is in fact a very primitive model: its concrete use requires
many fields (each molecule or ion a field!) and therefore
many phenomenological entries. 
Too small space-time scales must be avoided since a simple hard sphere
model is assumed for $V(r)$ at very small distances; still it copes
with a large part of physics.
A much more profound model would
already be available: electric charges carrying fields
interacting via electromagnetic field, that is QED. It is
used with strong schematizations in quantum optics and is a
well established subject in particle physics, which is
however generally biased by the old philosophy. When matter
inside the container is suitably isolated, in due time an
equilibrium state is prepared by spontaneous evolution of
the system; equilibrium states are then very well
described by  statistical operators
${\hat \varrho} (\beta,\mu)$ labeled by two parameters, temperature
$1/\beta$ and chemical potential $\mu$. These parameters
together with the geometry $\omega$ of the container provide
an objective and exhaustive characterization of the physical
system. The description given by the statistical operator is
essentially statistical: often the expectations of the
number of quanta in a normal mode $r$, i.e.,
$
        {\mbox{\rm Tr}}
        \left[
        {\hat a}_{r}^{\scriptscriptstyle\dagger}
        {\hat a}_{r}
        {\hat \varrho} (\beta,\mu)
        \right]
$, is exceedingly small (typically it approaches zero in the
limit of infinite volume of $\omega$, an impressive
exception is however ground state occupation in the case of
Bose Einstein condensation~\cite{Anderson}), indicating a strongly erratic
behavior of the microsystems in occupation of the huge set
of normal modes. If particles are introduced via
\textit{field
theory -- identical particle} correspondence, these particles
are entangled (this entanglement being just a key point for the
correspondence) leading to a variety of
low-temperature phenomena: only for high temperatures
entanglement can be forgotten and the picture of the  system
as composed by particles, each endowed with typical
dynamical properties, becomes useful.
By the quasi-local structure of the interaction term in
(\ref{1.1}) one is immediately led to the idea that settings
in a space region $\omega_I \subset \omega$ can be prepared
in which no interaction occurs,
i.e.,
$
        {\hat \psi}({\mbox{\bf x}},\sigma)
        {\hat \psi}({\mbox{\bf y}},\sigma')
        {\hat \varrho}=0
$,
$
{\mbox{\bf x}},{\mbox{\bf y}}
\in \omega_I
$,
so that the
dynamics of a single  microsystem can be exhibited. Then
also its interaction with other suitable systems can be
arranged, possibly interaction and entanglement with
other microsystems. Such an investigation is indeed
necessary to gain information on the {\it two-body} potential
$
        V(
        \left |
        {\bf x} - {\bf y}
        \right |
        )
$
in the energy density (\ref{1.1}), which represents the basic
phenomenological entry for the model we are considering:
similar problems arise also in QED, where the
electromagnetic structure of the microsystems is the basic
input.
From an experimental point of view all this has been
achieved to an astonishing level of performance:
a region
$\omega_I$ means vacuum technique, shielding devices, noise
control; systems feeding $\omega_I$ with possibly entangled
microsystems are now available sources;  systems that can be
macroscopically affected  by microsystems prepared in
$\omega_I$ are now available high efficiency detectors.
However such very satisfactory experimental situation has no
theoretical representation inside quantum field
theoretical description of systems. In fact such
description exists at present only for equilibrium systems,
while the situation we have now described is a highly
non-equilibrium situation: this is immediately clear if we think
about the extremely dishomogeneous mass distribution between 
$\omega_I$
and the rest of $\omega$. So there is at present a real gap inside
theoretical physics on the general subject of
non-equilibrium statistical mechanics. In \S~\ref{torun} some
proposal is presented to fill this gap; one expects that inside any
reasonable proposal a natural way should exist to characterize the
highly non-equilibrium situation in which a {\it source part} and a
{\it detecting part} of an isolated system can
interact through a directed microphysical channel.
However also in absence of this at present not yet available
theoretical setting, no problem did ever arise in the practical use of
quantum mechanics to describe what happens inside $\omega_I$.
Indeed there exists a large consistent theoretical frame including
quantum dynamics of microsystems (e.g., scattering processes) and
equilibrium statistical mechanics.
As was declared by Bell, quantum mechanics is valid {\it for all
practical purposes}~\cite{Bell}. But this is actually the point: quantum
mechanics is just the providential short cut through the
forementioned untilled ground. Clear statement of this role
of quantum mechanics dissipates, in our opinion, all that is
called problem on foundations of quantum mechanics, as we
will further discuss in the next sections.

\par  
\section{The role of quantum mechanics}
\label{surface}
\setcounter{equation}{0}
\par         
Let us assume that inside some region $\omega_I \subset\omega$ of the
system, interaction between the fields representing its
microphysical structure drops out, during a time interval
$[t_0,t_1]$: one has
        \begin{equation}
        \label{2.1}
        {\hat \psi}({\mbox{\bf x}},\sigma)
        {\hat \psi}({\mbox{\bf y}},\sigma)
        {\hat \varrho}_t=0,
        \qquad
        {\mbox{\bf x}},{\mbox{\bf y}}
        \in \omega_I,
        \qquad
        t\in [t_0,t_1].
        \end{equation}
This happens trivially if
        \[
        {\hat \psi}({\mbox{\bf y}},\sigma)
        {\hat \varrho}_t=0,
        \qquad
        {\mbox{\bf y}}
        \in \omega_I,
        \qquad
        t\in [t_0,t_1]
        \]
and in this case we shall say that in region $\omega_I$ one has
in this time interval a perfect vacuum.
Eq.~(\ref{2.1}) however also
holds if
${\hat \varrho}_t$ has the following more complex structure:
        \begin{equation}
        \label{2.3}
        {\hat \varrho}_t=
        \sum_{\sigma,\sigma'}
        \int_{\omega_I} d^3\!
        {\mbox{\bf y}}
        \,
        \int_{\omega_I} d^3\!
        {\mbox{\bf y}'}
        \,
        \Psi^{{\scriptscriptstyle (1)}}_t({\mbox{\bf y}},\sigma)
        {\hat \psi}^{\scriptscriptstyle\dagger}({\mbox{\bf y}},\sigma)
        {\hat \varrho}'_t
        {\hat \psi}({\mbox{\bf y}'},\sigma')
        \Psi^{{\scriptscriptstyle (1)}}_t{}^*({\mbox{\bf y}'},\sigma')
        \end{equation}
with
$
        {\hat \psi}({\mbox{\bf y}},\sigma)
        {\hat \varrho}'_t=0
$,
$
        {\mbox{\bf y}}
        \in \omega_I
$,
$
        t\in [t_0,t_1]
$ and
$
 \Psi^{{\scriptscriptstyle (1)}}_t({\mbox{\bf y}},\sigma)=0
$, 
$
{\mbox{\bf y}}\in\partial\omega_I
$.
One has:
        \begin{equation}
        \label{2.4}
        {\mbox{\rm Tr}}
        {\hat \varrho}_t
        =
        \sum_{\sigma}
        \int_{\omega_I} d^3\!
        {\mbox{\bf y}}
        \,
        {|
        \Psi^{{\scriptscriptstyle (1)}}_t({\mbox{\bf y}},\sigma)
        |}^2
        =1.
        \end{equation}
Using
$
[
        {\hat \psi}({\mbox{\bf y}},\sigma),
        {\hat \psi}^{\scriptscriptstyle\dagger}({\mbox{\bf
        y}'},\sigma')
]
=
\delta^3
({\mbox{\bf y}}-{\mbox{\bf y}'})
\delta{\sigma,\sigma'}
$ one can show that
${\hat \varrho}_t$ is a solution
of the Liouville -- von Neumann equation
$
d 
{\hat \varrho}_t
/ dt
=
-
{
i
\over
 \hbar
}
[
{\hat H}_t,{\hat \varrho}_t
]
$
with Hamiltonian (\ref{1.1}) whenever
${\hat \varrho}'_t$ is and if the function
$
        \Psi^{{\scriptscriptstyle (1)}}_t({\mbox{\bf y}},\sigma)
$
satisfies
a remarkable
equation.
This equation, if one considers points ${\mbox{\bf y}}$
at a distance from $\partial\omega_I$ larger than the range
of $V(r)$ and neglects an \textit{operator valued} surface term,
takes a system independent form and is precisely the Schr\"odinger
equation:
        \begin{equation}
        \label{2.5}
        i\hbar {\partial\over  \partial t}  
        \Psi^{{\scriptscriptstyle (1)}}_t({\mbox{\bf y}},\sigma)
        =-{
          \hbar^2  
        \over  
                 2m  
        }          \Delta_2  
        \Psi^{{\scriptscriptstyle (1)}}_t({\mbox{\bf y}},\sigma)
        +  
        U(
        {\mbox{\bf y}},t
        )  
        \Psi^{{\scriptscriptstyle (1)}}_t({\mbox{\bf y}},\sigma)
        \end{equation}
(near to the boundary $\partial\omega_I$ an effective additional
operator valued contribute to $U(
        {\mbox{\bf y}},t
        )  $ appears).
By (\ref{2.4}) the role of the Hilbert space
$
{\cal H}^{{\scriptscriptstyle (1)}}
=L^2(\omega_I)\otimes {\mbox{\bf C}}^{2s+1}
$
immediately appears as the natural setting for  (\ref{2.5})
and from now on we shall take over formalism and notation
developed by use of
$
{\cal H}^{{\scriptscriptstyle (1)}}
$
in quantum mechanics (bra, ket, observables).
An important generalization of the situation indicated by
(\ref{2.3}) occurs if ${\hat \varrho}_t$ has the form:
        \begin{equation}
        \label{2.6}
        {\hat \varrho}_t=
        \sum_{\sigma,\sigma'}
        \int_{\omega_I} d^3\!
        {\mbox{\bf y}}
        \,
        \int_{\omega_I} d^3\!
        {\mbox{\bf y}'}
        \,
        {\hat \psi}^{\scriptscriptstyle\dagger}({\mbox{\bf y}},\sigma)
        {\hat \varrho}'_t
        {\hat \psi}({\mbox{\bf y}'},\sigma')
        \langle
        {\mbox{\bf y}},\sigma
        |
        {\varrho}_t^{{\scriptscriptstyle (1)}}
        |
         {\mbox{\bf y}'},\sigma'
        \rangle
        ,
        \end{equation}
where ${\varrho}_t^{{\scriptscriptstyle (1)}}$
is a statistical operator on
$
{\cal H}^{{\scriptscriptstyle (1)}}
$,
such that
        \begin{equation}
        \label{2.7}
        {
        d
        {\varrho}_t^{{\scriptscriptstyle (1)}}
        \over
        dt
        }
        =
        -
        {
        i
        \over
         \hbar
        }
        [
        {H}_t^{{\scriptscriptstyle (1)}},
        {\varrho}_t^{{\scriptscriptstyle (1)}}
        ]
        ,
        \end{equation}
with
${H}_t^{{\scriptscriptstyle (1)}}$ the Hamilton operator by which
(\ref{2.5}) is written as
$
        i\hbar
        (
        {
        d
        \Psi^{{\scriptscriptstyle (1)}}
        /
        dt
        }
        )
        =
        {H}_t^{{\scriptscriptstyle (1)}}
        \Psi^{{\scriptscriptstyle (1)}}
        ;
$
still ${\hat \varrho}_t$ is a solution of the Liouville -- von
Neumann equation of the  system if ${\hat \varrho}'_t$ is and
if ${\varrho}_t^{{\scriptscriptstyle (1)}}$ satisfies (\ref{2.7}).
Representation (\ref{2.6}) becomes (\ref{2.3}) if
$
        {\varrho}_t^{{\scriptscriptstyle (1)}}
$
is a pure state:
        \begin{equation}
        \label{2.8}
        \langle
        {\mbox{\bf y}},\sigma
        |
        {\varrho}_t^{{\scriptscriptstyle (1)}}
        |
         {\mbox{\bf y}'},\sigma'
        \rangle
        =
        \Psi^{{\scriptscriptstyle (1)}}_t({\mbox{\bf y}},\sigma)
        \Psi^{{\scriptscriptstyle (1)}}_t{}^*({\mbox{\bf y}'},\sigma').
        \end{equation}
Let us further assume that for $t=t_0$,
$
        \Psi^{{\scriptscriptstyle (1)}}_t({\mbox{\bf y}},\sigma)
$
is practically different from zero only
in a region inside $\omega_I$ very small
at the space-scale of our macroscopic
description: then ${\hat \varrho}_{t_0}$ can be described
as the source of a microsystem localized in this region
at time $t_0$.
Let
${\hat A}$ be an observable of the system, typically a {\it
relevant variable} of a measured quantity (see \S~\ref{torun}) or
the spectral measure ${\hat E}^A(M)$ ($M$ being a Borel set), by which
expectations of ${\hat A}$ can be related to probability
distributions $p_t^A(M)$.
One has:
        \begin{eqnarray}
        \label{2.9}
        \langle
        {\hat A}
        \rangle_t
        &=&
        {\mbox{\rm Tr}}
        ({\hat A}{\hat \varrho}_t)
        \\
        &=&
        \sum_{\sigma,\sigma'}
        \int_{\omega_I} d^3\!
        {\mbox{\bf y}}
        \,
        \int_{\omega_I} d^3\!
        {\mbox{\bf y}'}
        \,
        \Psi^{{\scriptscriptstyle (1)}}_t({\mbox{\bf y}},\sigma)
        \langle
         {\mbox{\bf y}'},\sigma'
        |
        {A}_t^{{\scriptscriptstyle (1)}}
        |
        {\mbox{\bf y}},\sigma
        \rangle
        \Psi^{{\scriptscriptstyle (1)}}_t{}^*({\mbox{\bf y}'},\sigma')
        =
        \langle
        \Psi^{{\scriptscriptstyle (1)}}_t
        |
        A_t^{{\scriptscriptstyle (1)}}
        |
        \Psi^{{\scriptscriptstyle (1)}}_t
        \rangle
        \nonumber
        \\
        \label{2.10}
        p_t^A(M)
        &=&
        {\mbox{\rm Tr}}
        ({\hat E}^A(M){\hat \varrho}_t)
        =
        \langle
        \Psi^{{\scriptscriptstyle (1)}}_t
        |
        F_t^{{\scriptscriptstyle (1)}}{^A}(M)
        |
        \Psi^{{\scriptscriptstyle (1)}}_t
        \rangle
        \end{eqnarray}
where:
        \begin{equation}
        \label{2.11}
        \langle
         {\mbox{\bf y}'},\sigma'
        |
        {A}_t^{{\scriptscriptstyle (1)}}
        |
        {\mbox{\bf y}},\sigma
        \rangle
        =
        {\mbox{\rm Tr}}
        \left(
        {\hat A}
        {\hat \psi}^{\scriptscriptstyle\dagger}({\mbox{\bf y}},\sigma)
        {\hat \varrho}'_t
        {\hat \psi}({\mbox{\bf y}'},\sigma')
        \right)
        \end{equation}
        \begin{equation}
        \label{2.12}
        \langle
        {\mbox{\bf y}'},\sigma'
        |
        F_t^{{\scriptscriptstyle (1)}}{^A}(M)
        |
        {\mbox{\bf y}},\sigma
        \rangle
        =
        {\mbox{\rm Tr}}
        \left(
        {\hat E}^A(M)
        {\hat \psi}^{\scriptscriptstyle\dagger}({\mbox{\bf y}},\sigma)
        {\hat \varrho}'_t
        {\hat \psi}({\mbox{\bf y}'},\sigma')
        \right)
        .
        \end{equation}
More generally one has:
$
        \langle
        {\hat A}
        \rangle_t
        =
        {\mbox{\rm Tr}}_{{\cal H}^{(1)}}
        \left(
        A_t^{{\scriptscriptstyle (1)}}
        \varrho_t^{{\scriptscriptstyle (1)}}
        \right)
$,
$
        p_t^A(M)
        =
        {\mbox{\rm Tr}}_{{\cal H}^{{\scriptscriptstyle (1)}}}
        \left(
        F_t^{{\scriptscriptstyle (1)}}{}^A(M)
        \varrho_t^{{\scriptscriptstyle (1)}}
        \right)
$.
Expectations
$
        \langle
        {\hat A}
        \rangle_t
$
and probability  distributions
$
        p_t^A(M)
$
are related to symmetric operators
$A_t^{{\scriptscriptstyle (1)}}$ in
${\cal H}^{{\scriptscriptstyle (1)}}$ and to their p.o.v.~spectral
measures
$
        F_t^{{\scriptscriptstyle (1)}}{}^A(M)
$:
$
A_t^{{\scriptscriptstyle (1)}}
=
\int_{{\bf R}} \lambda dF_t^{{\scriptscriptstyle (1)}}{^A}
$.
Notice how naturally the well known Dirac's quantum
mechanical formalism arises in this context, that had
initially nothing to do with it.
Let us now discuss the basic relations (\ref{2.9}),
(\ref{2.10}): they provide the dynamics of the ({\it
macro})-system prepared as given by (\ref{2.3}) or
(\ref{2.6}). The dynamics consists in the time dependence of
$
        \Psi^{{\scriptscriptstyle (1)}}_t
$
(${\varrho}_t^{{\scriptscriptstyle (1)}}$) and of
$A_t^{{\scriptscriptstyle (1)}}$
and $F_t^{{\scriptscriptstyle (1)}}{^A}$. A basic difference now arises:
$
        \Psi^{{\scriptscriptstyle (1)}}_t
$
(${\varrho}_t^{{\scriptscriptstyle (1)}}$)
are solutions of a {\it universal} (as far as the
microphysical field theoretical model reaches) equation in
which only fundamental aspects of quasi-local
interactions enter. 
This is of course true in so far as the operator valued surface term
neglected in  (\ref{2.5}) and given explicitly by
\begin{eqnarray*}
\lefteqn{
{
          \hbar^2  
        \over  
                 2m          
}   
\left[
        \sum_{\sigma,\sigma'}
        \int_{\partial\omega_I} d^2\!
        {\mbox{\bf a}}
        \,
        \int_{\omega_I} d^3\!
        {\mbox{\bf y}'}
        \,
        {\bf n}_I \cdot
        \nabla
        \Psi^{{\scriptscriptstyle (1)}}_t({\mbox{\bf y}},\sigma)
        {\hat \psi}^{\scriptscriptstyle\dagger}({\mbox{\bf y}},\sigma)
        {\hat \varrho}'_t
        {\hat \psi}({\mbox{\bf y}'},\sigma')
        \Psi^{{\scriptscriptstyle (1)}}_t{}^*({\mbox{\bf y}'},\sigma')
\right.}
\\
&&
\left.
\hphantom{{
          \hbar^2  
        \over  
                 2m          
}   
\Delta_2    }
-
        \sum_{\sigma,\sigma'}
        \int_{\omega_I} d^3\!
        {\mbox{\bf y}}
        \,
        \int_{\partial\omega_I} d^2\!
        {\mbox{\bf a}'}
        \,
        {\bf n}_I \cdot
        \Psi^{{\scriptscriptstyle (1)}}_t({\mbox{\bf y}},\sigma)
        {\hat \psi}^{\scriptscriptstyle\dagger}({\mbox{\bf y}},\sigma)
        {\hat \varrho}'_t
        {\hat \psi}({\mbox{\bf y}'},\sigma')
        \nabla
        \Psi^{{\scriptscriptstyle (1)}}_t{}^*({\mbox{\bf y}'},\sigma')       
\right]
,  
\end{eqnarray*}
where ${\bf n}_I$ is the normal to the surface of $\omega_I$, can actually
be disregarded. The relevance of this contribution provides a control
mechanism for the feasibility of the considered description.
The ({\it
macro})-system only provides different initial conditions at
time $t_0$, i.e., different sources can be described, but
there is no other influence of ${\hat \varrho}_t$ on
$
        \Psi^{{\scriptscriptstyle (1)}}_t
$
(${\varrho}_t^{{\scriptscriptstyle (1)}}$).
The time dependence of
$A_t^{{\scriptscriptstyle (1)}}$
and $F_t^{{\scriptscriptstyle (1)}}{^A}$ is instead related to the time
evolution of ${\hat \varrho}'_t$,  that is to say to the irreversible
dynamics of the macroscopic background consisting of the
system, when the microsystem we are dealing with has been
cut out. Let us look closer to this time dependence. Taking
$
        {\hat \psi}({\mbox{\bf y}},\sigma)
        {\hat \varrho}'_t=0
$
into account
one has by (\ref{2.11}) and (\ref{2.12}):
        \begin{equation}
        \label{2.13}
        \langle
        {\mbox{\bf y}'},\sigma'
        |
        A_t^{{\scriptscriptstyle (1)}}
        |
        {\mbox{\bf y}},\sigma
        \rangle
        =
        \delta^3
        ({\mbox{\bf y}}-{\mbox{\bf y}'})
        \delta{\sigma,\sigma'}
        {\mbox{\rm Tr}}
        ({\hat A}{\hat \varrho}'_t)
        +
        {\mbox{\rm Tr}}
        [
        {\hat {\cal A}}
        (
        {\mbox{\bf y}'},\sigma';
        {\mbox{\bf y}},\sigma
        )
        {\hat \varrho}'_t
        ]
        \end{equation}
where:
        \begin{equation}
        \label{2.14}
        {\hat {\cal A}}
        (
        {\mbox{\bf y}'},\sigma';
        {\mbox{\bf y}},\sigma
        )
        =
        \frac 12
        \left(
        \left[
        {\hat \psi}({\mbox{\bf y}'},\sigma'),
        {\hat A}
        \right]
        {\hat \psi}^{\scriptscriptstyle\dagger}({\mbox{\bf y}},\sigma)
        +
        {\hat \psi}({\mbox{\bf y}'},\sigma')
        \left[
        {\hat A},
        {\hat \psi}^{\scriptscriptstyle\dagger}({\mbox{\bf y}},\sigma)
        \right]
        \right)
        .
        \end{equation}
By the splitting (\ref{2.13}),
$
A_t^{{\scriptscriptstyle (1)}}
=
{\bf 1}^{{\scriptscriptstyle (1)}}
        {\mbox{\rm Tr}} ({\hat A}{\hat \varrho}'_t)
        +
{\cal A}_t^{{\scriptscriptstyle (1)}}
$
or also
$
{\cal A}_t^{{\scriptscriptstyle (1)}}
=
\int_{{\bf R}}
(\lambda -
        {\mbox{\rm Tr}} ({\hat A}{\hat \varrho}'_t)
)
dF_t^{{\scriptscriptstyle (1)}}{^A}
$, a  {\it microsystem independent} dynamics of the system
is brought into evidence and can be considered separately.
On the contrary the operators
$
{\cal A}_t^{{\scriptscriptstyle (1)}}
$ provide a contribution to the dynamics of the microsystem,
arising from the background irreversible dynamics of the
system.
Now it becomes clear under what condition the macroscopic dynamics
of the system given by
$
        {\mbox{\rm Tr}}
        ({\hat A}{\hat \varrho}_t)
$,
$
        {\mbox{\rm Tr}}
        ({\hat E}^A(M){\hat \varrho}_t)
$
provides an insight into and is adequately described through the
dynamics of the microsystem, which is described by the {\it
universal} element
$
        \Psi^{{\scriptscriptstyle (1)}}_t
$
(${\varrho}_t^{{\scriptscriptstyle (1)}}$).
The additional time dependence induced by 
${\hat \varrho}'_t$
in
$A_t^{{\scriptscriptstyle (1)}}$
must be negligible during the time evolution of the
microsystem:
$A_t^{{\scriptscriptstyle (1)}}
\approx A^{{\scriptscriptstyle (1)}}
$
for 
$t\in[t_0,t_1]$. In this case the system
is not only the source but also becomes a
possible measuring device of the observable
$A^{{\scriptscriptstyle (1)}}$
of the microsystem and the p.o.v.~measure
$F^{{\scriptscriptstyle (1)}}{}^A$
provides its probability distribution. Of prominent
importance and simplicity are the projection valued measures
on ${\cal H}^{{\scriptscriptstyle (1)}}$:
if
$
F^{{\scriptscriptstyle (1)}}{}^A(M)=
(F^{{\scriptscriptstyle (1)}}{}^A(M))^2
$, then orthonormal resolutions of the identity
${\bf 1}^{{\scriptscriptstyle (1)}}$
are associated to measurements and the physics related to the
microsystem can be displayed by well-known quantum mechanics
using geometry of ${\cal H}^{{\scriptscriptstyle (1)}}$.
The main question is to
characterize ${\hat \varrho}_t$ so that
$A_t^{{\scriptscriptstyle (1)}}
\approx A^{{\scriptscriptstyle (1)}}
$
in correspondence to variables ${\hat A }$ of the system: is
this possible? This question cannot be solved explicitly at
the present stage; however just the experimental success of
quantum mechanics provides a phenomenological affirmative
answer. Not only, but typical basic features of quantum
mechanics indicate limitations in this assertion,
may be unexpected, at least if one still
endorses the old philosophy.
By the existence of non compatible observables in
${\cal H}^{{\scriptscriptstyle (1)}}$, e.g.,
position and momentum, one must
expect that for a given ${\hat \varrho}_t$ only for some
subset of variables
$A_t^{{\scriptscriptstyle (1)}}
\approx A^{{\scriptscriptstyle (1)}}
$
can hold: different ${\hat \varrho}_t$ must be taken if
$A^{{\scriptscriptstyle (1)}}$ is a sharp measurement
of position or if
$A^{{\scriptscriptstyle (1)}}$ is a sharp measurement of momentum;
similarly if
different components of spin are considered.
To enlighten what a  microsystem is, e.g., what
its spin is, related to a field
${\hat \psi}({\mbox{\bf x}},\sigma)$,
the dynamics of a
\textit{suitably large set of different (macro)-systems},
structured in terms of
${\hat \psi}({\mbox{\bf x}},\sigma)$ is necessary: thus this
micro-macrosystem relationship is deeply
different from the classical one. In our opinion
quantum mechanics is and must be felt puzzling and
paradoxical when just this point is missed. Let us now
return briefly to the simplest microsystem in our
model: if source and potential
${U}({\mbox{\bf x}},t)$ allow a classical limit the
naive concept of a particle with mass $m$ and
magnetic moment related to the spin $s$ can be
finally recovered in well known ways.
In the
general case the term {\it one quanton
microsystem} is more appropriated, as suggested in
a recent very pedagogical paper by
B.-G.~Englert~\cite{Englert}.
$
\Psi_t^{{\scriptscriptstyle (1)}}({\mbox{\bf y}},\sigma)
$
represents source and subsequent evolution,
$A^{{\scriptscriptstyle (1)}}$ an observable, all this can be embedded
in the dynamics of some suitable system ${\hat
\varrho}_t$. More complicated microsystems arise
in obvious way, taking for example
        \begin{eqnarray*}
        {\hat \varrho}_t
        &=&
        \sum_{\sigma_1,\sigma_2 \atop
        \sigma'_1,\sigma'_2}
        \int_{\omega_I} d^3\!
        {\mbox{\bf y}_1}
        \,
        \int_{\omega_I} d^3\!
        {\mbox{\bf y}_2}
        \,
        \int_{\omega_I} d^3\!
        {\mbox{\bf y}'_1}
        \,
        \int_{\omega_I} d^3\!
        {\mbox{\bf y}'_2}
        \,
        \nonumber
        \\
        &&
        \hphantom{      }
        {} \times
        {
        1
        \over
         2!
        }
        \Psi^{{\scriptscriptstyle (2)}}_t({\mbox{\bf
        y}_1},\sigma_1;{\mbox{\bf y}_2,\sigma_2)
         {\hat \psi}^{\scriptscriptstyle\dagger}({\mbox{\bf
        y}_1},\sigma_1)
        \psi}^{\scriptscriptstyle\dagger}({\mbox{\bf
        y}_2},\sigma_2)
        {\hat \varrho}'_t
        {\hat \psi}({\mbox{\bf y}'_2},\sigma'_2)
        {\hat \psi}({\mbox{\bf y}'_1},\sigma'_1)
        \Psi^{{\scriptscriptstyle (2)}}_t{}^*({\mbox{\bf
        y}'_1},\sigma'_1;{\mbox{\bf
        y}'_2},\sigma'_2)
        .
        \end{eqnarray*}
instead of equation (\ref{2.3}), one has a two quanton
microsystem described by the symmetric
($           
[
{\hat \psi}({\mbox{\bf y}_1},\sigma_1),
{\hat \psi}({\mbox{\bf
y}_2},\sigma_2)
]_- =0
$)
or the antisymmetric
($
[
{\hat \psi}({\mbox{\bf y}_1},\sigma_1),
{\hat \psi}({\mbox{\bf
y}_2},\sigma_2)
]_+ =0
$)
wave function
$
        \Psi^{{\scriptscriptstyle (2)}}_t({\mbox{\bf
        y}_1},\sigma_1;{\mbox{\bf y}_2},\sigma_2)
$: here entanglement between the two microsystems
is mandatory. By the dynamics of this ${\hat
\varrho}_t$ interaction between the two
microsystems can be investigated and the basic
entry
$
        V(
        \left |
        {\mbox{\bf x}} - {\mbox{\bf y}}
        \right |
        )
$
inside (\ref{1.1}), which also enters the
Schr\"odinger equation ruling
$
        \Psi^{{\scriptscriptstyle (2)}}_t
$, can be checked with high accuracy. Also to test
the extremely  simple structure of a total spin 0
system more than one system ${\hat \varrho}_t$ is
necessary. This is the root of the long debated
famous EPR paradox: when the spin of only one
of two spin 1/2 microsystems is measured, the
simple statistical operator
$
{
1
\over
2
}
{\bf 1}^{{\scriptscriptstyle (1)}}
$
in one particle spin space
${\bf C}^2$ arises. However measuring the spin of
the second particle this statistical operator  can
be demixed in different ways, according to which
component of the spin of the second particle is
measured: these are however measurements involving
different ${\hat \varrho}_t$. Demixtures to be
related to different choices of ${\hat \varrho}_t$
are called different {\it as if realities} or
different {\it blends} in the forementioned
Englert's paper. This is actually the most
reasonable point of view if quantum mechanics is
discussed as the {\it theory of microsystems}. A
very deep, unavoidably abstract and ponderous,
location of quantum mechanics as the description
of a microphysical channel inside a
phenomenological description of experimental
settings was given by Ludwig starting in the 60's
and collected in references~\cite{Ludwig}: in
this approach all usual paradoxes and puzzles were
already foreseen and reduced to reasonable
problems and also modern tools in measurement
theory like p.o.v.~measures, operations and
instruments were anticipated, even if with
different names. Our proposal points to a
realization of Ludwig's approach in the context
of a quantum field theoretical description of isolated
physical  systems.
The question
$A_t^{{\scriptscriptstyle (1)}}
\approx A^{{\scriptscriptstyle (1)}}
$, $t\in [t_0,t_1]$ has another far reaching
implication: one cannot expect that by right
choice of ${\hat \varrho}_t$ and of the
observables a strict time independence arises, i.e.,
$
dA_t^{{\scriptscriptstyle (1)}} / dt =0
$.
One must instead expect that a sufficiently weak
coupling to the irreversible dynamics of
${\hat \varrho}'_t$ exists, so that the universal
behavior of the microsystem is not too much
obscured by the system depending dynamics of
$A^{{\scriptscriptstyle (1)}}$.
The study of
$
        {\mbox{\rm Tr}}
        (
        {\hat {\cal A}}(
        {\mbox{\bf y}'},\sigma';
        {\mbox{\bf y}},\sigma
        )
        {\hat \varrho}'_t
        )
$
is a difficult many-body problem and since the
variables ${\mbox{\bf y}'},\sigma'$
and ${\mbox{\bf y}},\sigma$ are jointly involved
one can expect that just due to the weak coupling
situation the dynamics of the microsystem receives
non Hamiltonian corrections and that a dynamical
semigroup evolution for $\varrho^{{\scriptscriptstyle (1)}}$
arises.
One takes care in this way of the fact that the
universal behavior alone is a too strict
schematization, as it is already indicated by the
so called infrared problem of field theory. All
this is related to the debated problem of
decoherence: this concept arises now in a very
natural way. Obviously in our scheme any
modification of fundamental dynamics to add
decoherence would be 
superfluous: it is
automatically present if microsystems are
described, while objectivity of the description of
systems is granted by the structure of the theory,
as it will be shown in the next section.
The problem of preserving coherence in the physics of microsystems,
faced by experimentalists with modern high technology,
should have in our
opinion the theoretical counterpart in the forementioned study of 
$
        {\mbox{\rm Tr}}
        (
        {\hat {\cal A}}(
        {\mbox{\bf y}'},\sigma';
        {\mbox{\bf y}},\sigma
        )
        {\hat \varrho}'_t
        )
$
in
quantum field theory of non-equilibrium systems.

\par  
\section{Isolated non-equilibrium macrosystems}
\label{torun}
\setcounter{equation}{0}
\par
According to the point of view substantiated
in \S~\ref{lucrezio} and \S~\ref{surface}, which
establishes the foundations of  quantum theory on a suitable
theory of non-equilibrium quantum statistical mechanics,
we will sketch a proposal pointing in this direction, based
on some recent work~\cite{torun} and will give a brief account of it. 
In order to tackle the
problem it is quite natural to cope first with isolated
systems, this attitude being taken not only for the sake of
simplicity, but also in connection with the search for
objective state parameters, as we shall see later on.
The very concept of isolation is however far from trivial,
as it immediately appears taking into account the existence
of correlations and quantum mechanical entanglement, which
prevent any real physical system to be utterly isolated from
the rest of the world. This issue is strongly related to the
problem of decoherence (for a recent review see~\cite{Kiefer}) 
and its technical and
philosophical consequences. Despite these facts, the notion
of isolation is still meaningful and useful, provided it is
related to a specific time scale, so that only a suitable
subset of all possible observables is to be considered,
specifically those observables which are slowly varying on
this time scale. Relying on this restriction a physically
realizable  preparation  procedure should be conceivable,
which actually implements this effective isolation. On the
contrary if variables sensible to any time or energy scale
were considered, shielding from the influence of the
environment would be unfeasible. The set of fundamental fields
(e.g., associated to charged elementary particles as in QED,
or to molecules in a neutral continuum) and the choice of
relevant observables (e.g., hydrodynamic or kinetic
description of a massive continuum) determine the level of
description and therefore actually define the considered
macrosystem and its relevant time scale.
We are thus lead to look for \textit{subsets of relevant slow
variables}, and a natural choice are the densities of
conserved quantities. In fact, as stressed in non-equilibrium
statistical  mechanics (see for example~\cite{Zubarev}),
such densities, averaged with suitable probe functions, do
provide natural candidates for relevant, not too fast
changing observables. 
The basic structure which characterizes a suitable description of an
isolated system at a certain space-time scale is an adequate choice of
a subset ${\cal M}$ of \textit{relevant variables} $
        {\hat A}_j({\mbox{{\boldmath$\xi$}}})
$: for any realistic system ${\cal M}$ will not be invariant under time
evolution generated by the Hamiltonian of the model; this can be seen
as the root of the second principle of thermodynamics.

The set ${\cal A}_t$ of expectation
values of the relevant variables
$
        {\hat A}_j({\mbox{{\boldmath$\xi$}}})
$
at any time $t$ fixes a set
${\cal K}_{{\cal A}_t}$
of statistical operators, compatible with the given
expectations. For a  macrosystem one cannot generally expect
that
${\cal A}_t$ uniquely determines a statistical operator, so
that a further selection inside
${\cal K}_{{\cal A}_t}$
has to be carried out. This can be meaningfully accomplished
by maximizing the von Neumann entropy (supposed finite for
every element of
${\cal K}_{{\cal A}_t}$
), so that one is generally led to consider generalized
Gibbs states of the form
        \begin{equation}
        \label{a}
        {\hat w}[{\zeta(t)}]
        =
        {  
        \exp
        \left \{
        {-{
        \sum_j
        \int d {\mbox{{\boldmath$\xi$}}} \,
        {\zeta}_j({\mbox{{\boldmath$\xi$}}},t)
        {\hat A}_j({\mbox{{\boldmath$\xi$}}})
        }}  
        \right \}
        \over  
        Z
        \left[
        {\zeta}(t)
        \right]  
        }
        \equiv
        \exp
        \left \{
        {
        -\zeta_0(t) {\hat {\bf 1}}
        -{
        \sum_j
        \int d {\mbox{{\boldmath$\xi$}}} \,
        {\zeta}_j({\mbox{{\boldmath$\xi$}}},t)
        {\hat A}_j({\mbox{{\boldmath$\xi$}}})
        }}  
        \right \}
        \end{equation}
where
$
        \zeta_0(t)=\log
        Z
        \left[
        {\zeta}(t)
        \right]
$,
$
        Z
        \left[
        {\zeta}(t)
        \right]  
        =
        {\mbox{{\rm Tr}}} \,
        {  
        \exp
        \left \{
        {-{
        \sum_j
        \int d {\mbox{{\boldmath$\xi$}}} \,
        {\zeta}_j({\mbox{{\boldmath$\xi$}}},t)
        {\hat A}_j({\mbox{{\boldmath$\xi$}}})
        }}  
        \right \}
        }
$
being the partition function of the system at time $t$.
The maximization of
$-k{\mbox{{\rm Tr}}} \, {\hat \sigma}
\log {\hat \sigma}
$,
$
{\hat \sigma}\in
{\cal K}_{{\cal A}_t}
$,
leads to the appearance of the
Lagrange parameters
${\zeta}_j({\mbox{{\boldmath$\xi$}}},t)$
linked to the mean values at time $t$ of the relevant
observables. These classical parameters
represent a generalization of 
temperature and chemical potential that we meet in the usual applications
of equilibrium  statistical mechanics; they characterize the
statistical collection and can be naturally taken as
properties of each member of the  statistical collection,
thus assuming an objective meaning. 
These can be considered as the \textit{beables}, so called in contrast to
\textit{observables} by Bell, who claimed physics should be founded on the
former rather than on the latter ones.
This is substantiated by
phenomenological evidences, for example in the description
of macrosystems inside mechanics of continua, where the
velocity field
$
        {\mbox{\bf v}}({\mbox{{\bf x}}},t)
$
(related to the reference frame in which the continuum is at
rest) can be endowed with an objective meaning for each
individual system, together with the fact that
$
        {\mbox{\bf v}}({\mbox{{\bf x}}},t)
$
is given by the ratio between the mean values of momentum
density and mass density in the statistical collection. 
This distinguished role of mean
values in leading to objective features of the statistical
collection, reflecting the fact that the measurement of
these relevant observables does indeed provide a negligible
influence on their expectation values, is confirmed in the
framework of continuous measurement theory~\cite{continue}.
There one can check that in the limit in which the coupling
to the measuring apparatus, responsible for the continuous
observation and the removal of the system's isolation,
vanishes, the trajectories for the mean values still remain
meaningful, while second or higher order momenta of the
probability distribution diverge.
Entropy of the system is defined as a function of 
$
\left \{
\zeta(t)
\right \}
$
by
\begin{displaymath}
-k
{\mbox{{\rm Tr}}} \,
{\hat w}[{\zeta(t)}]   \log
{\hat w}[{\zeta(t)}]
.
\end{displaymath}

\par
In order to give the dynamics of the macrosystem for times
$t>t_0$ (supposing the isolation effective with respect to
the relevant variables begins at time $t_0$)
we have to give a recipe for the determination of the
statistical operator ${\hat \varrho}_t$ of the macrosystem, with
which to calculate the mean values and therefore determine
the state parameters. A
straightforward and naive approach would be to take
        \begin{equation}
        \label{1t}
        {\hat \varrho}_{t_0}  =
        {\hat w}[{\zeta(t_0)}]
        \end{equation}
with
$
        {\hat w}[{\zeta(t_0)}]
$
given by (\ref{a}) and exploiting for later times the
unitary evolution, we would have:
        \begin{equation}
        \label{c}
        \hat\varrho_t=\hat U(t-t_0)\hat
        w [{\zeta}(t_0)]
        \hat U^{\scriptscriptstyle\dagger}(t-t_0)=
        \exp
        \left \{
        {
        -\zeta_0(t_0) {\hat {\bf 1}}
        -{
        \sum_j
        \int d {\mbox{{\boldmath$\xi$}}} \,
        {\zeta}_j({\mbox{{\boldmath$\xi$}}},t_0)
        {\hat A}_j({\mbox{{\boldmath$\xi$}}},-(t-t_0))
        }}  
        \right \}
        \,
        .
        \end{equation}
The choice (\ref{1t}), corresponding to the standpoint of
{\it information thermodynamics}, amounts to considering the
history up to the time point $t_0$ completely negligible,
while this is no more true for later times, as can be seen
from  (\ref{c}). A more realistic viewpoint consists in
taking ${\hat \varrho}_t$ as the representative of the spontaneous
time evolution the system has undergone from time $t_0$ to
time $t$ and of a suitable  preparation  procedure operated
in the finite time interval
$[T,t_0]$, controlling and measuring the system before
enforcing isolation at time $t_0$. In this perspective the
following quantities should be considered in assigning
${\hat \varrho}_{t_0}$ as a result of a concrete preparation
procedure:
the expectations of the relevant variables at sharp
time points $T$ and $t_0$; the previous history in the time interval
$[T,t_0]$ controlled by measurements of variables
$
\int_T^{t_0} dt' \,
{\hat A}_j({\mbox{{\boldmath$\xi$}}},t') h_\alpha (t')
$
(with $h_\alpha (t')$
suitable test functions, e.g., $h_\alpha (t)=\cos
\omega_\alpha t$). Together with
the densities
${\hat A}_j({\mbox{{\boldmath$\xi$}}},t)$ one should also
consider the corresponding currents
${\hat {\bf J}}_j({\mbox{{\boldmath$\xi$}}})$ related to
them by conservation equations
of the form
$
        {\dot {{\hat A}_j}} ({\mbox{{\boldmath$\xi$}}},t)
        =
        - \nabla \cdot {\hat {\bf J}}_j({\mbox{{\boldmath$\xi$}}},t)
$,
where time dependence is given in Heisenberg picture
$
        {\hat A}_j({\mbox{{\boldmath$\xi$}}},t)
        =
        e^{+{{
        i
        \over
         \hbar
        }}{\hat H}t}
        {\hat A}_j({\mbox{{\boldmath$\xi$}}})
        e^{-{{
        i
        \over
         \hbar
        }}{\hat H}t}
$.
In the end one obtains
        \begin{eqnarray}
        \label{2t}
        {\hat \varrho}_{t_0}
        &=&
        \exp
        \left \{
        -\sum_j
        \int d {\mbox{{\boldmath$\xi$}}} \,
        \gamma_j ({\mbox{{\boldmath$\xi$}}},{t_0})
        {\hat A}_j({\mbox{{\boldmath$\xi$}}})
        +
        \sum_{j \alpha}
        \int d {\mbox{{\boldmath$\xi$}}}  \,
        \gamma_{j\alpha} ({\mbox{{\boldmath$\xi$}}})
        \int_T^{t_0}  dt' \,
        {\hat A}_j({\mbox{{\boldmath$\xi$}}},-(t_0 - t'))
        h_{j\alpha}(t')
        \right.
        \nonumber
        \\
        &&
        \hphantom{
        \exp
        \left \{
        \right.
        }
        +
        \sum_{j \alpha}
        \int d {\mbox{{\boldmath$\xi$}}} \,
        {\mbox{{\boldmath $\gamma$}}}_{j\alpha}
        ({\mbox{{\boldmath$\xi$}}})
        \cdot
        \int_T^{t_0} dt' \,
        {\hat {\bf J}}_j({\mbox{{\boldmath$\xi$}}},-(t_0 - t'))
        h_{j\alpha}(t')
        \nonumber
        \\
        &&
        \hphantom{
        \exp
        \left \{
        \right.
        }
        \left.
        -\sum_j
        \int d {\mbox{{\boldmath$\xi$}}} \,
        \gamma_j ({\mbox{{\boldmath$\xi$}}},T)
        {\hat A}_j({\mbox{{\boldmath$\xi$}}},-(t_0 -T))
        \right \}
        .
        \end{eqnarray}
A crucial step is now to assume that
        \begin{equation}
        \label{3t}
        \gamma_j ({\mbox{{\boldmath$\xi$}}},{t_0})
        =
        \zeta_j ({\mbox{{\boldmath$\xi$}}},{t_0})
        \end{equation}
so as to stress the distinguished role of the expectations
of the relevant observables
${\hat A}_j({\mbox{{\boldmath$\xi$}}},t)$
at time $t_0$,
$
\left \{
\zeta(t_0)
\right \}
$
being the parameters characterizing the macroscopic state.
Eq.~(\ref{3t}) will hold at least for some suitable
preparation  procedures.
In this way a time arrow is introduced, because of the
asymmetry between
$
        \gamma_j ({\mbox{{\boldmath$\xi$}}},{t_0})
$
and
$
        \gamma_j ({\mbox{{\boldmath$\xi$}}},{T})
$.
One has therefore, considering time evolution  of the
isolated system up to time $t$:
        \begin{eqnarray}
        \label{4t}
        {\hat \varrho}_t
        =
        e^{-{{
        i
        \over
         \hbar
        }}{\hat H}(t-t_0)}
        {\hat \varrho}_{t_0}
        e^{+{{
        i
        \over
         \hbar
        }}{\hat H}(t-t_0)}
        &=& 
        \exp
        \left \{
        {}  
        -\zeta_0 (t) {\hat {\bf 1}}
        -\sum_j
        \int d {\mbox{{\boldmath$\xi$}}} \,
        \zeta_j ({\mbox{{\boldmath$\xi$}}},{t_0})
        {\hat A}_j({\mbox{{\boldmath$\xi$}}},-(t-t_0))
        \right.
        \nonumber
        \\
        &&
        \hphantom{
        \exp
        \left \{
        \right.
        }
        +
        \sum_{j \alpha}
        \int d {\mbox{{\boldmath$\xi$}}} \,
        \gamma_{j\alpha} ({\mbox{{\boldmath$\xi$}}})
        \int_T^{t_0} dt'\,
        {\hat A}_j({\mbox{{\boldmath$\xi$}}},-(t- t'))
        h_{j\alpha}(t')
        \nonumber
        \\
        &&
        \hphantom{
        \exp
        \left \{
        \right.
        }
        +
        \sum_{j \alpha}
        \int d {\mbox{{\boldmath$\xi$}}} \,
        {\mbox{{\boldmath $\gamma$}}}_{j\alpha}
        ({\mbox{{\boldmath$\xi$}}})
        \cdot
        \int_T^{t_0} dt'\,
        {\hat {\bf J}}_j({\mbox{{\boldmath$\xi$}}},-(t - t'))
        h_{j\alpha}(t')
        \nonumber
        \\
        &&
        \hphantom{
        \exp
        \left \{
        \right.
        }
        \left.
        -
        \sum_j
        \int d {\mbox{{\boldmath$\xi$}}} \,
        \gamma_j ({\mbox{{\boldmath$\xi$}}},T)
        {\hat A}_j({\mbox{{\boldmath$\xi$}}},-(t -T))
        \right \}
        .
        \end{eqnarray}
Exploiting this expression
one can determine the macrostate at time $t$, using the
expectations
\begin{equation}
\label{valori medi}
 \langle
{\hat A}_j({\mbox{{\boldmath$\xi$}}})
\rangle_t
=
{\mbox{{\rm Tr}}} \,
(
{\hat A}_j({\mbox{{\boldmath$\xi$}}})
{\hat \varrho}_t
) 
\end{equation}
to determine the parameters
$
\left \{
\zeta(t)
\right \}
$
in
$
        {\hat w}[{\zeta(t)}]
$.
Eq.~(\ref{4t}) and (\ref{valori medi}) give the objective dynamics of
the system. The existence of $
\left \{
\zeta(t)
\right \}
$ is granted by the mathematical structure. An immense complexity is
hidden in the Heisenberg picture of the operators inside (\ref{4t})
for any realistic system, despite the apparently simple formulas.
Using the functions $\zeta_j({\mbox{{\boldmath$\xi$}}},t)$
we can rewrite
$
        \zeta_j({\mbox{{\boldmath$\xi$}}},t_0)
        {\hat A}_j({\mbox{{\boldmath$\xi$}}},-(t-t_0))
$ in the following way:
        \begin{eqnarray*}
        &&
        \zeta_j({\mbox{{\boldmath$\xi$}}},t_0)
        {\hat A}_j({\mbox{{\boldmath$\xi$}}},-(t-t_0))
        =
        \zeta_j({\mbox{{\boldmath$\xi$}}},t)
        {\hat A}_j({\mbox{{\boldmath$\xi$}}})
        -
        \int_{t_0}^t
        dt' \,
        {
        d
        \over
         dt'
        }
        \left[
        \zeta_j({\mbox{{\boldmath$\xi$}}},t')
        {\hat A}_j({\mbox{{\boldmath$\xi$}}},-(t-t'))
        \right]
        \\
        &&
        =
        \zeta_j({\mbox{{\boldmath$\xi$}}},t)
        {\hat A}_j({\mbox{{\boldmath$\xi$}}})
        -
        \int_{t_0}^t
        dt' \,
        {\dot{\zeta}}_j({\mbox{{\boldmath$\xi$}}},t')
        {\hat A}_j({\mbox{{\boldmath$\xi$}}},-(t-t'))
        -
        \int_{t_0}^t
        dt' \,
        \zeta_j({\mbox{{\boldmath$\xi$}}},t')
        {\dot {{\hat A}_j} }({\mbox{{\boldmath$\xi$}}},-(t-t'))
        \nonumber
        \\
        &&
        =
        \zeta_j({\mbox{{\boldmath$\xi$}}},t)
        {\hat A}_j({\mbox{{\boldmath$\xi$}}})
        -
        \int_{t_0}^t
        dt' \,
        {\dot{\zeta}}_j({\mbox{{\boldmath$\xi$}}},t')
        {\hat A}_j({\mbox{{\boldmath$\xi$}}},-(t-t'))
        +
        \int_{t_0}^t
        dt' \,
        \zeta_j({\mbox{{\boldmath$\xi$}}},t')
         \nabla \cdot  {\hat {\bf J}}_j({\mbox{{\boldmath$\xi$}}},-(t-t'))
        ,
        \nonumber
        \end{eqnarray*}
so that we can rewrite (\ref{4t}) in the form
        \begin{eqnarray}
        \label{6t}
        {\hat \varrho}_t
        \!\!
        &=&
        \!\!
        \exp
        \left \{
        {}
        -\zeta_0(t)  {\hat {\bf 1}}
        -\sum_j
        \int d {\mbox{{\boldmath$\xi$}}} \,
        \zeta_j ({\mbox{{\boldmath$\xi$}}},{t})
        {\hat A}_j({\mbox{{\boldmath$\xi$}}})
        +
        \sum_j
        \int_{t_0}^t dt'
        \int d {\mbox{{\boldmath$\xi$}}} \,
        {\dot{\zeta}}_j ({\mbox{{\boldmath$\xi$}}},{t'})
        {\hat A}_j({\mbox{{\boldmath$\xi$}}},-(t- t'))
        \right.
        \nonumber
        \\
        \!\!
        &&
        \hphantom{
        \exp
        \left \{
        \right.
        }
        {}
        +
        \sum_{j \alpha}
        \int^{t_0}_T dt'
        \int d {\mbox{{\boldmath$\xi$}}} \,
        \gamma_{j\alpha} ({\mbox{{\boldmath$\xi$}}})
        {\hat A}_j({\mbox{{\boldmath$\xi$}}},-(t- t'))
        h_{j\alpha}(t')
        \nonumber
        \\
        \!\!
        &&
        \hphantom{
        \exp
        \left \{
        \right.
        }
        {}
        -
        \sum_j
        \int_{t_0}^t dt'
        \int d {\mbox{{\boldmath$\xi$}}} \,
        {{\zeta}}_j ({\mbox{{\boldmath$\xi$}}},{t'})
        { \nabla \cdot
        {\hat {\bf J}}}_j({\mbox{{\boldmath$\xi$}}},-(t- t'))
        \nonumber
        \\
        \!\!
        &&
        \hphantom{
        \exp
        \left \{
        \right.
        }
        {}
        +
        \sum_{j \alpha}
        \int_T^{t_0} dt'
        \int d {\mbox{{\boldmath$\xi$}}} \,
        {\mbox{{\boldmath $\gamma$}}}_{j\alpha}
        ({\mbox{{\boldmath$\xi$}}})
        \cdot
        {\hat {\bf J}}_j({\mbox{{\boldmath$\xi$}}},-(t - t'))
        h_{j\alpha}(t')
        \nonumber
        \\
        \!\!
        &&
        \hphantom{
        \exp
        \left \{
        \right.
        }
        {}
        \left.
        -
        \sum_j
        \int d {\mbox{{\boldmath$\xi$}}} \,
        \gamma_j ({\mbox{{\boldmath$\xi$}}},T)
        {\hat A}_j({\mbox{{\boldmath$\xi$}}},-(t -T))
        \right \}
        .
        \end{eqnarray}
Comparing ${\hat \varrho}_t$ given by (\ref{6t}) with
${\hat \varrho}_{t_0}$ given by (\ref{2t}) one sees
that the basic structure is preserved:
${\hat \varrho}_t$ accounts for a preparation  procedure
terminating at time $t$, replacing $t_0$, the initial
parameters
$
\left \{
\zeta(t_0)
\right \}
$
being
replaced by
$
\left \{
\zeta(t)
\right \}
$. The
contribution referring to the past history extends now from
$T$ to $t$ and a new part is displayed, related to the time
interval $[t_0,t]$. In place of the parameters
$
\sum_{\alpha}
\gamma_{j\alpha}({\mbox{{\boldmath$\xi$}}}) h_{j\alpha}(t')
$
relative to the preparation  procedure in the time
interval $[T,t_0]$, now the parameters ${\dot
{\zeta}}_j({\mbox{{\boldmath$\xi$}}},t)$ appear, in place of the
term
$
        \sum_{j \alpha}
        {\mbox{{\boldmath $\gamma$}}}_{j\alpha}
        ({\mbox{{\boldmath$\xi$}}})
        \cdot
        {\hat {\bf J}}_j({\mbox{{\boldmath$\xi$}}},-(t - t'))
        h_{j\alpha}(t')
$
one deals with
$
        - \sum_j
        {{\zeta}}_j ({\mbox{{\boldmath$\xi$}}},{t'})
        { \nabla \cdot
        {\hat {\bf J}}}_j({\mbox{{\boldmath$\xi$}}},-(t- t'))
$.
This internal consistency gives the  \textit{justification for the
assumptions} (\ref{2t}),  (\ref{3t}) for the
${\hat \varrho}_{t_0}$ assigned to a suitable
preparation  procedure of a macrosystem. 
The structure of the obtained $\hat\varrho_t$ looks very
similar to the
{\it {non-equilibrium statistical operator}} proposed by
Zubarev~\cite{Zubarev}. A main difference has however to be
pointed out:
the choice $T\to-\infty$ in
Zubarev's approach means that no preparation time interval
is taken into account.
The limit $T\to -\infty$ presupposes a thermodynamic limit,
and therefore the idealization of infinite systems. This is
a very useful technical procedure in order to get rid of
boundary conditions, but should be avoided in a fundamental
approach, finite-sized  systems being often the realistic,
experimentally testable realizations of interesting physical
systems (e.g., the recent and outstanding example of
Bose Einstein condensation).
Nevertheless the huge set of practical applications of Zubarev's
formalism also fit in the present scheme.

Let us give the part
$        \sum_j
        \int d {\mbox{{\boldmath$\xi$}}} \,
        \zeta_j ({\mbox{{\boldmath$\xi$}}},{t})
        {\hat A}_j({\mbox{{\boldmath$\xi$}}})
$
for the case of hydrodynamical description of a continuum in terms of
the typical parameters 
$
        \beta({\mbox{\bf x}},t)
$,
$
        \mu({\mbox{\bf x}},t)
$,
$
{\mbox{\bf
v}}({\mbox{{\bf x}}},t)
$:
\begin{displaymath}
        {\int}_{\omega} d^3\!
        {\mbox{\bf x}}
        \,
        \beta({\mbox{\bf x}})
        \left[
        {\hat e}_o ({\mbox{\bf x}},
        {\mbox{\bf v}}({\mbox{{\bf x}}},t))
        +
        {
        \mu ({\mbox{\bf x}})
        \over
        m
        }
        {\hat \rho}({\mbox{\bf x}})
       \right]
\end{displaymath}
with
        \begin{eqnarray*}
        {\hat e} ({\mbox{\bf x}})
        &=&
        {\hat e}_o ({\mbox{\bf x}},
        {\mbox{\bf v}}({\mbox{{\bf x}}},t))
        +
        {\mbox{\bf v}}({\mbox{{\bf x}}},t)
        \cdot
        {\hat {\mbox{\bf p}}}_o ({\mbox{\bf x}},
        {\mbox{\bf v}}({\mbox{{\bf x}}},t))
        +
        {\scriptstyle {\frac 12}}
        {\mbox{\bf v}}^2 ({\mbox{{\bf x}}},t)
        {\hat \rho}({\mbox{\bf x}})
        \\
        {\hat {\mbox{\bf p}}} ({\mbox{\bf x}})
        &=&
        {\hat {\mbox{\bf p}}}_o ({\mbox{\bf x}},
        {\mbox{\bf v}}({\mbox{{\bf x}}},t))
        +
        {\mbox{\bf v}} ({\mbox{{\bf x}}},t)
        {\hat \rho}({\mbox{\bf x}})
        .
        \nonumber
        \end{eqnarray*}
Notice that
$
        \beta({\mbox{\bf x}},t)
$
is related to the energy density ${\hat e}_o ({\mbox{\bf x}})$  in the
local rest frame, obtained by replacing $
        \mp
        i\hbar
        \nabla
$
with
$
        \mp
        i\hbar
        \nabla
        - m
        {\mbox{\bf v}}({\mbox{{\bf x}}},t)
$
in the expressions for energy and momentum density.

Let us indicate in a sketchy way how one obtains from the general
equation the most simple form of quantum dynamics of a macroscopic
system. We shall call this situation, which already copes with so
called linear non-equilibrium thermodynamics, \textit{simple dynamics}.
Looking at (\ref{6t}) it is quite natural to exploit a
perturbative expansion around the first two  contributions
in the argument of the exponential, linked to normalization
and mean value of the relevant observables. 
Indeed by  (\ref{3t}) the first term is already all that is needed for
the dynamics of 
$
        {\hat A}_j({\mbox{{\boldmath$\xi$}}}) \in {\cal M}
$.
This can be
neatly done in terms of the so-called {\it cumulant
expansion}~\cite{Robin}, which allows to express the first
order perturbation in terms of two point {\it Kubo
correlation functions}.
Setting
$
        {\hat W}
        =
        e^{{\hat A}}/
        {
        {\mbox{{\rm Tr}}} \,
        e^{{\hat A}}
        }
$ one has
        \begin{eqnarray}
        \label{7t}
        {
        {\mbox{{\rm Tr}}} \,
        {\hat C}
        e^{{\hat A}+{\hat B}}
        \over
        {\mbox{{\rm Tr}}} \,
        e^{{\hat A}+{\hat B}}
        }
        &=&
        {
        {\mbox{{\rm Tr}}} \,
        {\hat C} {\hat W}
        }
        +
        {
        {\mbox{{\rm Tr}}} \,
        {\hat C}
        \int_0^1 du \,
        e^{u{\hat A}}
        {\hat B}
        e^{-u{\hat A}} {\hat W}
        }
        -
        {
        {\mbox{{\rm Tr}}} \,
        {\hat C} {\hat W}
        }
        {
        {\mbox{{\rm Tr}}} \,
        {\hat B} {\hat W}
        }
        +
        \ldots
        \nonumber
        \\
        &\equiv&
        {
        {\mbox{{\rm Tr}}} \,
        {\hat C} {\hat W}
        }
        +
        \mbox{\boldmath $\langle$}
        {\hat C},{\hat B}
        \mbox{\boldmath $\rangle$}_{{\hat W}}
        +
        \ldots
        \end{eqnarray}
where the Kubo correlation function with respect to the
statistical operator ${\hat W}$ has been implicitly
introduced.
By the very definition of macrostate one may write
        \begin{displaymath}
        {
        d
        \over
         dt
        }
        {\mbox{{\rm Tr}}} \,
        (
        {\hat A}_j({\mbox{{\boldmath$\xi$}}})
        {\hat \varrho}_t
        )
        =
        {
        d
        \over
         dt
        }
        {\mbox{{\rm Tr}}} \,
        (
        {\hat A}_j({\mbox{{\boldmath$\xi$}}})
        {{\hat w}[{\zeta{(t)}}]}
        )
        =
        -
        \sum_{l}
        \int d {\mbox{{\boldmath$\xi$}}}' \,
        \mbox{\boldmath $\langle$}
        {\hat A}_j({\mbox{{\boldmath$\xi$}}})   ,
        {\hat A}_{l}({\mbox{{\boldmath$\xi$}}}')
        \mbox{\boldmath $\rangle$}_{{\hat w}[{\zeta{(t)}}]}
        {\dot {\zeta}}_{l}({\mbox{{\boldmath$\xi$}}}',t),
        \end{displaymath}
with
$
{\hat w}[\zeta(t)]
=
        \exp
        \left \{
        {
        -\zeta_0(t) {\hat {\bf 1}}
        -{
        \sum_j
        \int d {\mbox{{\boldmath$\xi$}}} \,
        {\zeta}_j({\mbox{{\boldmath$\xi$}}},t)
        {\hat A}_j({\mbox{{\boldmath$\xi$}}})
        }}  
        \right \}
$ as in (\ref{a}), so that, exploiting the Liouville -- von
Neumann equation and the cumulant expansion (\ref{7t}) one
comes to an evolution equation for the parameters
$
\left \{
\zeta(t)
\right \}
$
in the form:
        \begin{eqnarray}
        \label{9t}
        \lefteqn{
        -  
        \sum_{l}
        \int d {\mbox{{\boldmath$\xi$}}}' \,
        \mbox{\boldmath $\langle$}
        {\hat A}_j({\mbox{{\boldmath$\xi$}}})   ,
        {\hat A}_{l}({\mbox{{\boldmath$\xi$}}}')
        \mbox{\boldmath $\rangle$}_{{\hat w}[{\zeta{(t)}}]}
        {\dot {\zeta}}_{l}({\mbox{{\boldmath$\xi$}}}',t)
        =
        }
        \\
        &&
        {\mbox{{\rm Tr}}} \,
        \left(
        {
        i
        \over
         \hbar
        }
        [
        {\hat H},
        {\hat A}_j({\mbox{{\boldmath$\xi$}}})
        ]
        {\hat w}[{\zeta{(t)}}]
        \right)
        +
        \int_T^t dt'\,
        \mbox{\boldmath $\langle$}
        {
        i
        \over
         \hbar
        }
        [
        {\hat H},
        {\hat A}_j({\mbox{{\boldmath$\xi$}}})
        ]
        ,
        {\hat {\cal S}}(t')
        \mbox{\boldmath $\rangle$}_{{\hat w}[{\zeta{(t)}}]}
        \nonumber
        \\
        &&
        -
        \sum_{l}
        \int d {\mbox{{\boldmath$\xi$}}}' \,
        \mbox{\boldmath $\langle$}
        {
        i
        \over
         \hbar
        }
        [
        {\hat H},
        {\hat A}_j({\mbox{{\boldmath$\xi$}}})
        ]
        ,
        {\hat A}_{l}({\mbox{{\boldmath$\xi$}}}',-(t-T))
        \mbox{\boldmath $\rangle$}_{{\hat w}[{\zeta{(t)}}]}
        { {\gamma}}_{l}({\mbox{{\boldmath$\xi$}}}',T)
        + \ldots
        \>
        .
        \nonumber
        \end{eqnarray}
The term
$
        {\hat {\cal S}}(t')
$, keeping track of the previous history together with
$
        { {\gamma}}_{l}({\mbox{{\boldmath$\xi$}}}',T)
$, has a different expression for the two different time
intervals
$[T,t_0]$ and
$[t_0,t]$    relative to the preparation time and the
spontaneous time evolution respectively. The precise
structure of
$
        {\hat {\cal S}}(t')
$
can be read from (\ref{6t}) and is given by
        \begin{displaymath}
        \left \{  
        \begin{array}{lll}
        \sum_{j \alpha}
        \int d {\mbox{{\boldmath$\xi$}}}' \,
        [
        \gamma_{j\alpha} ({\mbox{{\boldmath$\xi$}}}')
        {\hat A}_j({\mbox{{\boldmath$\xi$}}}',-(t- t'))
        +
        {\mbox{{\boldmath $\gamma$}}}_{j\alpha}
        ({\mbox{{\boldmath$\xi$}}}')
        \cdot
        {\hat {\bf J}}_j({\mbox{{\boldmath$\xi$}}}',-(t - t'))
        ]
        h_{j\alpha}(t')
        &
        \quad
        T\leq t'\leq t_0
        \\  
        \vphantom{alza}
        \\  
        \sum_j
        \int d {\mbox{{\boldmath$\xi$}}}'\,
        [
        {\dot{\zeta}}_j ({\mbox{{\boldmath$\xi$}}}',{t'})
        {\hat A}_j({\mbox{{\boldmath$\xi$}}}',-(t- t'))
        -
        {{\zeta}}_j ({\mbox{{\boldmath$\xi$}}}',{t'})
        { \nabla \cdot
        {\hat {\bf J}}}_j({\mbox{{\boldmath$\xi$}}}',-(t- t'))
        ]
        &
        \quad
        t_0\leq t'\leq t
        \end{array}  
        \right.
        ,
        \end{displaymath}
so that (\ref{9t}) is in fact an integrodifferential
equation for
$\zeta_j({\mbox{{\boldmath$\xi$}}},t)$.
It appears that memory
of the macrostate through the whole interval $T\leq t'\leq
t$ is expressed in the first order approximation with
respect to (\ref{7t}) by the following two point Kubo
correlation functions:
        \[
        \mbox{\boldmath $\langle$}
        {
        i
        \over
         \hbar
        }
        [
        {\hat H},
        {\hat A}_j({\mbox{{\boldmath$\xi$}}})
        ]
        ,
        {\hat A}_{l}({\mbox{{\boldmath$\xi$}}}',-(t-t'))
        \mbox{\boldmath $\rangle$}_{{\hat w}[{\zeta{(t)}}]}
        ,
        \quad
        \mbox{\boldmath $\langle$}
        {
        i
        \over
         \hbar
        }
        [
        {\hat H},
        {\hat A}_j({\mbox{{\boldmath$\xi$}}})
        ]
        ,
        {\hat {\bf J}}_l({\mbox{{\boldmath$\xi$}}}',-(t-t'))
        \mbox{\boldmath $\rangle$}_{{\hat w}[{\zeta{(t)}}]}
        ,
        \quad
        T \leq t'\leq t
        ,
        \]
higher order correlation functions appearing in
higher order approximations, which can consistently be
neglected in most applications of
non-equilibrium thermodynamics.
A crucial point, in order to actually
calculate the dynamics of a complex system, is linked to \textit{the
decaying behavior of correlation functions}.
This is here not achieved by means of a thermodynamic limit,
the finite size of the system leading to a quasiperiodical
behavior of correlation functions, but depends on their
integration with respect to time and configuration space
with suitable smooth functions, as provided by
$h_{j\alpha}(t')$ and the parameters
$\zeta_j({\mbox{{\boldmath$\xi$}}},t')$ themselves, which
should be
homogeneous enough, provided one has done a
suitable
choice of the space-time variation scale of the relevant
observables.
Calling $\tau$ the characteristic correlation decay time,
the integration over history in (\ref{9t}) can be restricted
to the interval
$[t-\tau,t]$, so that the integrodifferential equations
governing the dynamics depend only the state parameters
and not on the preparation parameters, provided
$t_0 -T> \tau$, thus acquiring a universal character.
In this situation increase of entropy defined by
$
-k
{\mbox{{\rm Tr}}} \,
{\hat w}[{\zeta(t)}]   \log
{\hat w}[{\zeta(t)}]
$
and approach to an equilibrium Gibbs state can be shown.
A particularly simple but relevant situation arises if in
fact $\tau$ settles a typical time scale of the
slow dynamics
of the relevant observables over which the parameters
$
\left \{
\zeta(t)
\right \}
$
are
practically constant.
In this case the role of the statistical operator
${\hat \varrho}_t$ is played by the Gibbs state
${\hat w}[{\zeta{(t)}}]$ and one obtains that the expectations of
the relevant observables  are driven by a
dynamical semigroup of mappings acting on the
linear space generated by these observables.
In this way the formalism of the {\it master-equation}
appears again, though in a new light. One is not looking for
the time evolution of some coarse grained  statistical
operator,
but rather works in Heisenberg picture on a subset of
relevant slow variables.

\par  
\section{Memory and microsystems}
\label{memoria}
\setcounter{equation}{0}
\par     
Inside the general scheme presented
in \S~\ref{torun} a dramatic simplification
was introduced by the concept of \textit{simple dynamics} and by
relying on \textit{decaying behavior} of Kubo correlation
functions. These correlation functions become then the driving element
of dynamical evolution: in \textit{simple dynamics} two-point
correlations already do the job. Together with typical field
theoretical locality this leads to locality of macroscopic dynamics,
loosely speaking, evolution at a space-time point depends on local
macroscopic state in a short time interval before. In a relativistic
context this point becomes still sharper due to
micro-causality. However memory loss is only a general possibility
related to an interplay between the quasiperiodical behavior of
correlation functions and the more or  less smooth space-time
behavior of macroscopic state parameters: it is not a
necessary feature. If you consider a system composed by the source of a
microsystem, some intermediate device lingering with the microsystem
for some time $\Delta t$ and a detector, you have immediate
evidence of a system completely outside the previous description.
Assume that dynamics described by 
${\hat \varrho}_t$ in (\ref{6t}) was \textit{normal} (i.e., memory
confined inside very small time interval $\tau$) until time ${\bar t}$.
Let us now pose a seed of non normal behavior in some future
time interval and describe in not too technical way how this can be
done.
In the history part:
        \begin{displaymath}
        \int_{{\bar t} - \tau}^{{\bar t}} dt'\,
        {\hat {\cal S}}(t')
        \end{displaymath}
at the exponent of the expression of ${\hat \varrho}_t$ given by
(\ref{6t}), we cut out (setting e.g.
$
{\zeta}(t')=
{\zeta}_n(t')+{\tilde\zeta}(t')
)$ 
a part 
$
 {\hat {\cal S}}_{{\bar t} }(t')
$
which possibly will not simply decay:
$
 {\hat {\cal S}}(t')= {\hat {\cal S}}_{n}(t')
 + {\hat {\cal S}}_{{\bar t} }(t')
$.
Then ${\hat \varrho}_{{\bar t}}$ has a structure of the form
        \begin{equation}
        \label{4.1}
        {\hat \varrho}_{{\bar t}}
        =
        \exp
        \left\{
        -
        {\hat {\cal F}}({{\bar t}})
        +
        \int_{{\bar t} - \tau}^{{\bar t}} dt'\,
        {\hat {\cal S}}_{n}(t')
        +
        \int_{{\bar t} - \tau}^{{\bar t}} dt'\,
        {\hat {\cal S}}_{{\bar t} }(t')
        \right\}
        .
        \end{equation}
This positive operator can be represented as the normal statistical
operator
${\hat \varrho}_{{\bar t},n}$
(consisting of (\ref{4.1})
with normal history part
$
{\hat {\cal S}}_{n}(t')
$,
properly normalized)
with a suitable correction in the following form:
        \begin{equation}
        \label{4.2}
        {\hat \varrho}_{\bar t}
        =
        \lambda
        \left[
        {\hat {\bf 1}}
        +
        \int_{{\bar t} - \tau}^{{\bar t}} dt'\,
        {\hat {\mbox{\~{S}}}}_{{\bar t}}(t')
        \right]
        {\hat \varrho}_{{\bar t},n}
        \left[
        {\hat {\bf 1}}
        +
        \int_{{\bar t} - \tau}^{{\bar t}} dt'\,
        {\hat {\mbox{\~{S}}}}^{\scriptscriptstyle\dagger}_{{\bar t}}(t')
        \right],
        \end{equation}
where 
$
        {\hat {\mbox{\~{S}}}}_{{\bar t}}(t')
$
is essentially constructed with
$
        {\hat {\cal S}}_{{\bar t}}(t')
$.
Assume now that, due to some strong dishomogeneity of the classical
parameters in 
${\hat \varrho}_{\bar t,n}$,
there is a region 
$\omega_I$
such that
        \begin{equation}
        \label{4.3}
        {\hat \psi}({\mbox{\bf y}},\sigma)
        {\hat \varrho}_{\bar t,n}
        \approx 0
        \qquad
        {\mbox{\bf y}}
        \in \omega_I;
        \end{equation}
then the simplest representation of
$
        {\hat {\mbox{\~{S}}}}_{{\bar t}}(t')
$
is:
        \begin{displaymath}
        {\hat {\mbox{\~{S}}}}_{{\bar t}}(t')
        {\hat \varrho}_{\bar t,n}
        =
        \sum_{\sigma}
        \int_{\omega_I} d^3\!
        {\mbox{\bf y}}
        \,
        {\hat \psi}^{\scriptscriptstyle\dagger}({\mbox{\bf y}},\sigma)
        {\hat A}
        ({\mbox{\bf y}},\sigma,t')
        {\hat \varrho}_{\bar t,n}
        \end{displaymath}
where
$
        {\hat A}
        ({\mbox{\bf y}},\sigma,t')
$
acts as a destruction operator only in a region
$\omega_S$ outside
$\omega_I$ ($\omega_S \cap \omega_I= \emptyset $).
In general  by (\ref{4.3})
a strongly dishomogeneous (i.e., non-equilibrium) situation of the
system is stated. Typically one can  think of a \textit{source}
located in $\omega_S$
which feeds $\omega_I$ with one microsystem, taking it in some
complicated way
from region $\omega_S$.
These correction factors in (\ref{4.2})
give a decaying contribution to
expectations of observables which are influenced by the structure
of ${\hat \varrho}_{\bar t,n}$ inside $\omega_S$:
typically the dynamics of the source itself is
\textit{normal}.
Assume however
that there are observables
not influenced by the structure of ${\hat \varrho}_{\bar t,n}$
inside $\omega_S$. This is obviously a very
particular situation: consider an observable ${\hat B}$
of the system that
is shielded against the region $\omega_S$, except for the
microsystem; one can
formalize this by
$
        {\mbox{\rm Tr}}
        \left(
        {\hat B}
        \hat U(t-{\bar t})
        {\hat \varrho}_{\bar t,n}
        \hat U^{\scriptscriptstyle\dagger}(t-{\bar t})
        \right)
        \approx
        0
$.
With this condition one can see that just
the part of  (\ref{4.2})
bilinear in
$
        {\hat {\mbox{\~{S}}}}_{{\bar t}}(t')
$
becomes important and provides a
contribution of the form:
        \[
        {\mbox{\rm Tr}}
        (
        {\hat B}
        {\hat \varrho}_{t}
        )
        =
        (1-\lambda)
        {\mbox{\rm Tr}}
        \left(
        {\hat B}
        \hat U(t-{\bar t})
        {\hat \varrho}_{\bar t,a}
        \hat U^{\scriptscriptstyle\dagger}(t-{\bar t})
        \right),
        \qquad
        0<\lambda<1,
        \]
        \begin{equation}
        \label{4.5}
        {\hat \varrho}_{\bar t,a}
        =
        \sum_{\sigma,\sigma'}
        \int_{\omega_I} d^3\!
        {\mbox{\bf y}}
        \,
        \int_{\omega_I} d^3\!
        {\mbox{\bf y}'}
        \,
        {\hat \psi}^{\scriptscriptstyle\dagger}({\mbox{\bf y}},\sigma)
        {\hat \varrho}_{\bar t,n}
        {\hat \psi}({\mbox{\bf y}'},\sigma')
        \langle
        {\mbox{\bf y}},\sigma
        |
        {\varrho}_{{\bar t}}^{{\scriptscriptstyle (1)}}
        |
         {\mbox{\bf y}'},\sigma'
        \rangle
        \end{equation}
where
        \[
        \langle
        {\mbox{\bf y}},\sigma
        |
        {\varrho}_{{\bar t}}^{{\scriptscriptstyle (1)}}
        |
         {\mbox{\bf y}'},\sigma'
        \rangle
        =
        {
        {\mbox{\rm Tr}}
        \left(
        {\hat A}
        ({\mbox{\bf y}},\sigma)
        {\hat \varrho}_{\bar t,n}
        {\hat A}^{\scriptscriptstyle\dagger}
        ({\mbox{\bf y}'},\sigma')
        \right)
        \over
        \sum_{\sigma}
        \int_{\omega_I} d^3\!
        {\mbox{\bf y}}
        \,
        {\mbox{\rm Tr}}
        \left(
        {\hat A}
        ({\mbox{\bf y}},\sigma)
        {\hat \varrho}_{\bar t,n}
        {\hat A}^{\scriptscriptstyle\dagger}
        ({\mbox{\bf y}'},\sigma')
        \right)
        },
        \qquad
        {\hat A}
        ({\mbox{\bf y}},\sigma)
        =
        \int_{{\bar t} - \tau}^{{\bar t}} dt'\,
        {\hat A}
        ({\mbox{\bf y}},\sigma,t')
        .
        \]
The study of the time evolution of the statistical operator
$
        \hat U(t-{\bar t})
        {\hat \varrho}_{\bar t,a}
        \hat U^{\scriptscriptstyle\dagger}(t-{\bar t})
$
was just
the subject of \S~\ref{surface}, where this operator was
simply called
${\hat \varrho}_{t}$, (while
${\hat \varrho}'_{t}=
        \hat U(t-{\bar t})
        {\hat \varrho}_{\bar t,n}
        \hat U^{\scriptscriptstyle\dagger}(t-{\bar t})
$).
The mixture parameter $\lambda$ has the obvious
interpretation as probability
that region $\omega_S$ does not become active as a
source of a microsystem. The
\textit{anomalous} dynamics of the system around time ${\bar t}$,
consisting in the structure
$
{\hat \varrho}_{\bar t}=
\lambda{\hat \varrho}_{\bar t,n}
+
(1-\lambda)
{\hat \varrho}_{\bar t,a}
$,
is also indicated by the fact that one cannot longer state
that entropy increases around time ${\bar t}$.
One could say that an \textit{event} can happen with
probability ($1-\lambda$) in the time
interval $[{\bar t}-\tau,{\bar t}\,]$,
by which a new structure, the microsystem, is seeded, which
propagates inside $\omega_I$
and thus produces a more complex dynamics which
conserves memory of what was going on inside $\omega_S$
in the time interval $[{\bar t}-\tau,{\bar t}\,]$, to
be detected by observables ${\hat B}$ at time $t$; the value
$|t- {\bar t}|$
depending on how
successful fight against decoherence was (see \S~\ref{surface}).

\par  
\section{Conclusions and outlook}
\label{end}
\setcounter{equation}{0}
\par         
In \S~\ref{torun} and \S~\ref{memoria} we have indicated a possible
way of tackling quantum
field theory to construct a strongly non-equilibrium  dynamics. A
natural opening seems to appear indicating microsystems as
representatives of fundamental and universal local properties of
matter and as systems storing the physics of \textit{source} parts and
transmitting this memory to detecting parts. In this way entanglement
and interaction with other microsystems can be achieved giving an
outlook over more complex memory storing structures and
\textit{anomalous} (that is non entropy increasing) dynamics. What was
described in \S~\ref{memoria} happening around time ${{\bar t}}$
could be called an
\textit{event} and some relation with ideas recently expressed by
Haag~\cite{Haag} can be expected. Fortunately enough quantum mechanics
exists independently: one has only to \textit{guess} properly what
element of
$
{\cal H}^{{\scriptscriptstyle (1)}}
$
represents concrete sources and detectors. If this is skillfully done
and also decoherence is taken into account by some correct
weak-coupling to the macroscopic environment, all this works very well
in any practical application. No fundamental problem appears but it
should become clear that \textit{Dirac's book} quantum
mechanics is an
exceedingly smart short pass across this quantum field theory of
non-equilibrium systems.
If one goes hiking on a smart short pass across a nontrivial landscape,
believing to be on a main road, to encounter puzzling situations is a
common experience. So one does not wonder that quantum mechanics is
so puzzling and mysterious that Bell claimed that it contains the
seeds of its own dissolution.

\vskip 10pt
{\bf Acknowledgments}
\par
B.~V. thanks the Alexander von
Humboldt-Stiftung for financial support.

\par  
  

\vskip 20pt  
\begin{thebibliography}{99} 

\bibitem{Anderson}
{M.~Anderson~{\it et al.}},
{Science}
{\bf 269},
198
(1995).
\par

\bibitem{Bell}
{J.~Bell},
{Phys. World}
{\bf 2},
33
(1990).
\par

\bibitem{Englert}
{B.-G.~Englert},
{Z.~Naturforsch.}
{\bf 54a},
11
(1999).
\par

\bibitem{Ludwig}
{G.~Ludwig},
{\it An Axiomatic Basis for Quantum Mechanics}
(Springer, Berlin, 1985);
{\it Foundations of Quantum
Mechanics}
({Springer}, {Berlin}, 1983){}.
\par

\bibitem{torun}
L.~Lanz, O.~Melsheimer and B.~Vacchini,
to appear on {\it Rep.~Math.~Phys.}
\par

\bibitem{Kiefer}
{D. Giulini~{\it et al.}},
{\it Decoherence and the Appearance of a Classical World in
Quantum Theory}
(Springer, Berlin, 1996).
\par

\bibitem{Zubarev}
{D.~N.~Zubarev},  
{\it Non-equilibrium statistical thermodynamics,}  
(1974, Consultant Bureau, New York);
{D. N. Zubarev, V. Morozov and G. Roepke},
{\it Statistical mechanics of non-equilibrium processes},
(Akademie-Verlag, Berlin, 1996).
\par

\bibitem{continue}
{A.~Barchielli, L.~Lanz, G.~M.~Prosperi},
{Nuovo Cimento}
{\bf 72B},
{79}
(1982);
{Found. Phys.}
{\bf 13},
779
(1983).
\par

\bibitem{Robin}
{W.~A.~Robin},
{J.~Phys.~A}
{\bf 23},
{2065}
({1990}).
\par

\bibitem{Haag}
R.~Haag,
Commun.~Math.~Phys.
{\bf 180},
733
(1996);
in {\it Lecture Notes in
Physics}, Vol. 517, Ph.~Blanchard and A.~Jadczyk eds.
({Springer}, {Berlin}, 1999).

\end{thebibliography}
\end{document}